\documentclass[aps,prx,reprint,superscriptaddress,longbibliography,
               nofootinbib]{revtex4-2}

\usepackage[utf8]{inputenc}
\usepackage[T1]{fontenc}
\usepackage{amsmath,amssymb,bm}
\usepackage{graphicx}
\usepackage{xcolor}
\usepackage{dsfont}
\usepackage{hyperref}
\hypersetup{colorlinks=true, linkcolor=blue!60!black,
            citecolor=blue!60!black, urlcolor=blue!60!black}

\usepackage{cleveref}
\crefname{figure}{Fig.}{Figs.}
\crefname{table}{Table}{Tables}
\crefname{equation}{Eq.}{Eqs.}
\crefname{section}{Sec.}{Secs.}
\crefname{subsection}{Sec.}{Secs.}
\crefname{subsubsection}{Sec.}{Secs.}
\crefname{appendix}{Appendix}{Appendices}

\newcommand{\ket}[1]{|#1\rangle}
\newcommand{\bra}[1]{\langle#1|}

\newcommand{\HU}{H_U}                      
\newcommand{\HK}{H_K}                      
\newcommand{\HOm}{H_\Omega}                
\newcommand{\Om}{\Omega}                   
\newcommand{\Omc}{\Omega_c}                
\newcommand{\dgap}[1]{\Delta_{#1}}         
\newcommand{\tee}{\gamma_{\rm \scriptscriptstyle E}}   

\begin{document}

\title{Topological phase transition in a symmetric blockade structure}

\author{Johannes M\"ogerle}
\affiliation{Institute for Theoretical Physics III and Center for
  Integrated Quantum Science and Technology, University of Stuttgart,
  70550 Stuttgart, Germany}
\author{Simon Fell}
\affiliation{Institute for Theoretical Physics III and Center for
  Integrated Quantum Science and Technology, University of Stuttgart,
  70550 Stuttgart, Germany}
\author{Tobias F. Maier}
\affiliation{Institute for Theoretical Physics III and Center for
  Integrated Quantum Science and Technology, University of Stuttgart,
  70550 Stuttgart, Germany}

\author{Ali Otaifi}
\affiliation{Department of Physics and Arnold Sommerfeld Center for Theoretical Physics (ASC),
  Ludwig Maximilian University of Munich, 80333 Munich, Germany}
\affiliation{Munich Center for Quantum Science and Technology (MCQST), 80799 Munich, Germany}
\author{Lode Pollet}
\affiliation{Department of Physics and Arnold Sommerfeld Center for Theoretical Physics (ASC),
  Ludwig Maximilian University of Munich, 80333 Munich, Germany}
\affiliation{Munich Center for Quantum Science and Technology (MCQST), 80799 Munich, Germany}

\author{Nicolai Lang}
\affiliation{Institute for Theoretical Physics III and Center for
  Integrated Quantum Science and Technology, University of Stuttgart,
  70550 Stuttgart, Germany}
\author{Hans Peter B\"uchler}
\affiliation{Institute for Theoretical Physics III and Center for
  Integrated Quantum Science and Technology, University of Stuttgart,
  70550 Stuttgart, Germany}

\date{\today}

\begin{abstract}
  We present a numerical study of a blockade Hamiltonian with a local $\mathbb{Z}_2$ symmetry.
  For such models, the existence of a topologically ordered ground state has been proven rigorously for weak Rabi driving [\href{https://doi.org/10.1103/dtlf-2q82}{T.~F.~Maier~\emph{et~al.}, PRX Quantum \textbf{6}, 030340 (2025)}].
  However, the phase diagram beyond weak driving is still an open question.
  We perform large-scale infinite-size density matrix renormalization group (iDMRG) simulations on a cylinder and map out the full phase diagram of the model.
  We find a single continuous phase transition between the topologically ordered phase and a trivial paramagnetic phase, and confirm numerically the stability
  of the topologically ordered phase up to Rabi drives comparable to the characteristic energy scale of the system.
  Furthermore, we compute the excitation gaps and show that the charge gap remains open across the phase transition and charges become confined in the trivial phase.
  On the other hand, the gap in the symmetric sector closes at the transition, consistent with the condensation of flux excitations.
\end{abstract}

\maketitle

\section{Introduction}
\label{sec:intro}

A remarkable property of quantum matter is the emergence of macroscopic phenomena from simple
microscopic interactions. One of the most exotic forms of such an emergent behavior are topologically ordered
phases, which are characterized by long-range entanglement and anyonic quasi-particle excitations; a phenomenon
going beyond the concept of spontaneous symmetry breaking~\cite{Wen2017,ChenGuWen2010}.
Paradigmatic examples are the toric code and more general quantum
doubles~\cite{Kitaev2003}, as well as string-net models~\cite{LevinWen2005},
all of which are exactly solvable fixed points of topological phases. Topological orders have potentially interesting applications as quantum memories and for
topological quantum computation~\cite{Dennis2002, Nayak2008, Terhal2015}. However, most of the well-understood models require local
multi-body interactions which do not naturally appear in real materials; a remarkable exception being
Kitaev's honeycomb model \cite{Kitaev2006Anyons}. This motivates the search for alternative
routes to realize topologically ordered states of quantum matter in experimentally accessible systems based on
simple two-body interactions. Recently it has been shown, based on a mathematically rigorous analysis, that blockade Hamiltonians
can give rise to topologically ordered phases~\cite{Maier2025TopologicalOrder, Buechler2025QuantumDoubles}. These models can be viewed
as transverse-field Ising models in the presence of local longitudinal fields and with intricate designs of strong, antiferromagnetic Ising interactions; these models are motivated by the blockade interaction between Rydberg atoms.
Here, we present a numerical study of a closely related model, and map out its full phase diagram as well as
the behavior of the excitation gap.

Topologically ordered phases naturally emerge in fractional quantum Hall systems~\cite{Tsui1982, Laughlin1983} from
the combination of high magnetic fields and strong Coulomb interactions, and more recently also in moir\'e fractional
Chern insulators in the absence of a magnetic field~\cite{Cai2023, Park2023}. Lately, much effort has been devoted to the search for quantum
phases with toric code topological order in the controlled environment of quantum simulators. Early focus in
condensed matter physics has been on Josephson junction arrays modeling dimer models with a spin liquid
ground state~\cite{Ioffe2002}, and on frustrated magnets~\cite{Jackeli2009, Takagi2019}
where strong spin--orbit coupling generates exchange interactions corresponding to Kitaev's honeycomb model~\cite{Kitaev2006Anyons}.
However, conclusive evidence for a quantum spin liquid ground state in these materials is still missing~\cite{Broholm2020}.
In synthetic quantum systems, on the other hand, first proposals focused on the Kitaev model with optical lattices~\cite{Duan2003}
and polar molecules~\cite{Micheli2006}, or fractional quantum Hall states with rotating gases~\cite{Cooper2008}, as well as
bosonic fractional Chern insulators with polar molecules~\cite{Yao2013} and Rydberg atoms~\cite{Weber2022}.
In particular the proposals for the realization of a spin liquid ground state in Rydberg atom arrays
with blockade interactions~\cite{Samajdar2021,Verresen2021,Tarabunga2022,VerresenVishwanath2022} have attracted considerable interest. However, first
experimentally observed signatures~\cite{Semeghini2021} turned out to be dynamical artifacts of the preparation
protocol~\cite{Giudici2022, Sahay2022}, while on other platforms only small instances of topological states
have been realized~\cite{Clark2020,Leonard2023}. Recently, a promising alternative route was identified where
local symmetries are built into the design of the Hamiltonian~\cite{Chamon2020,ChamonKerman2021, Maier2025TopologicalOrder, Buechler2025QuantumDoubles}.
For such models, the existence of a topologically ordered ground state has been proven rigorously for weak
Rabi driving~\cite{Maier2025TopologicalOrder, Buechler2025QuantumDoubles}. However, to understand the behavior of these topological
phases beyond the weak driving limit, it is mandatory to map out the full phase diagram by numerics and study
the stability of the topological phase under experimentally realistic conditions.

\begin{figure*}[tbh]
  \centering
  \includegraphics{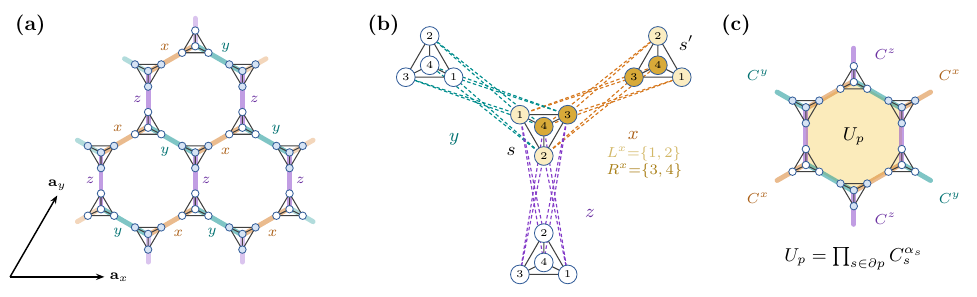}
  \caption{
    \textit{Setup of the blockade Hamiltonian.}
    (a) Honeycomb lattice with four two-level systems per site.
    On each site, the vertices describing the two-level systems are fully connected and account
    for the blockade interaction $V$ amongst them. By contrast, the blockade interaction
    between adjacent sites depends on the link direction $\alpha \in \{x,y,z\}$.
    (b) Detailed view of the blockade interaction between adjacent sites. The dashed lines describe the graph for
    blockade interaction of strength $U$ with the following construction: for each link $\alpha$,
    the four two-level systems on a site $s$ are partitioned into two classes $L_s^{\alpha}$ and $R_s^{\alpha}$. Then, a vertex
    $i$ on site $s$ interacts with a vertex $j$ on the neighboring site $s'$ along the link $\alpha$ if and only if they are in
    different classes, i.e., $i \in L_s^{\alpha}$ and $j \in R_{s'}^{\alpha}$ or vice versa. This construction is illustrated for
    an $x$-link. Note that we have placed the vertices on the sites such that the graph exhibits the full point group symmetry of
    the honeycomb lattice with the vertices $i \in R^\alpha$ on a line along the $\alpha$ link.
    (c) The graph constructed in this way exhibits a local group of graph automorphisms. On each site around a plaquette $p$,
    the permutation $C_s^{\alpha}$ exchanges the two vertices in the partition $L_s^{\alpha}$ and the two vertices in the
    partition $R_s^{\alpha}$. The product of these permutations around a plaquette, with the label $\alpha$ being the link pointing
    outward of the plaquette, defines such a local graph automorphism. The corresponding operator on the Hilbert space of
    the two-level systems then defines a local $\mathbb{Z}_2$ symmetry denoted by $U_p$.
  }
  \label{fig:model}
\end{figure*}

In this work, we present a numerical study of the phase diagram of a model based on blockade interactions with a local symmetry.
The model is closely related to the fully symmetric blockade structure introduced in Ref.~\cite{Maier2025TopologicalOrder}, but requires fewer atoms
per plaquette and is therefore more amenable to numerical simulations. The analysis is based on large-scale
infinite-size density matrix renormalization group (iDMRG) simulations~\cite{White1992DMRG,Schollwoeck2011DMRG} on a cylinder
using the open source software package \texttt{TeNPy}~\cite{Hauschild2024TeNPy}.
We derive the phase diagram as a function of the Rabi drive for fixed interaction strength, and find a single
continuous phase transition between the $\mathbb{Z}_2$ topological phase with toric code topological order and a
trivial paramagnetic phase. The transition is characterized by a closing of the excitation gap in the symmetric sector,
a jump in the topological entanglement entropy~\cite{Kitaev2006TEE,Levin2006Detecting,Jiang2012Identifying, Cincio2013Characterizing, Zhang2012MES},
and a vanishing expectation value of a Wilson loop operator. Remarkably, we find that
the charge gap -- the gap between different symmetry sectors -- remains open across the phase transition. Finally,
we derive the full phase diagram and the behavior of the critical Rabi drive as a function of the interaction strength. We apply
a reduced basis method~\cite{brehmer2023rb} to efficiently scan the phase diagram over a wide range of parameters.
This analysis complements the mathematically rigorous analysis of
Refs.~\cite{Maier2025TopologicalOrder, Buechler2025QuantumDoubles}, demonstrates that the topological phase spans a large parameter regime,
and opens the door for experimental realizations of topological orders by means of fully symmetric blockade structures
in Rydberg atom arrays or on alternative platforms.

\section{Model}
\label{sec:model}

The model derives from the blockade structures introduced in Refs.~\cite{Maier2025TopologicalOrder,Buechler2025QuantumDoubles}, namely
structures of two-level systems arranged in space in the presence of a uniform transverse field $\Omega$ and local detunings
$\Delta_i$, where the excited states of the two-level systems interact via a repulsive interaction. Such blockade structures
can be viewed as transverse-field Ising models in the presence of local longitudinal fields, where the complexity of the
system is encoded in the graph describing the antiferromagnetic Ising interactions. This class of Hamiltonians is most conveniently
described by a vertex-weighted graph, with each vertex corresponding to a two-level system, the weights accounting for the local
detunings, and edges corresponding to the blockade interaction between the two-level systems.

\subsection{Blockade Hamiltonian}

We focus on a setup on a honeycomb lattice with four two-level systems on each site of the honeycomb
lattice, see \cref{fig:model}(a); in the following, $|0\rangle_i$ denotes the ground state and $|1\rangle_i$ the
excited state of the two-level system $i$. On each site, these four two-level systems interact with a blockade
interaction $V$ amongst each other. We are interested in the limit $V\rightarrow \infty$ such that each site
is in full blockade and allows only for a single excitation. In addition, we assume a uniform detuning $\Delta_i\equiv\Delta$
for all two-level systems, such that the classical degenerate ground state configurations are given by a single
excitation per site of the honeycomb lattice. Furthermore, there is a blockade interaction $U$ between some vertices
on adjacent sites of the honeycomb lattice which follows the pattern shown by the graph in \cref{fig:model}(b);
the structure of these blockades is discussed below. While the detuning and the blockade interaction stabilize purely
classical ground state configurations, the non-trivial quantum dynamics is introduced by a homogeneous Rabi drive $\Om$.
Then, the Hamiltonian of the system is given by
\begin{align}
  H \;=\; \HOm + H_\Delta + H_V + \HU
  \label{eq:H}
\end{align}
where the coherent Rabi coupling takes the form
\begin{align}
  \HOm = \Om \sum_{i} \bigl(  \ket{1}\!\bra{0}_i + \ket{0}\!\bra{1}_i \bigr)\,,
  \label{eq:HOmega-def}
\end{align}
and the term accounting for the detuning is conveniently written as
\begin{align}
  H_\Delta = \Delta \sum_{s} \bigl( \mathds{1} - n_s \bigr)
  \quad\text{with}\quad
  n_s = \sum_{i \in s} n_i\,.
  \label{eq:HDelta}
\end{align}
Here, $s$ labels the sites of the honeycomb lattice with four vertices each, and $n_i = \ket{1}\!\bra{1}_i$ is the excitation number
operator for each two-level system $i$ on these sites. In the following, we fix the value of
the detuning $\Delta$ and express all other parameters (such as $\Omega$ and $U$) in units of $\Delta$.
The blockade interaction between two-level systems on the same site of the honeycomb lattice reduces to
\begin{align}
  H_V = \frac{V}{2} \sum_s \sum_{i\neq j \in s} n_i\, n_j\,.
  \label{eq:HV}
\end{align}
To describe the blockade interaction $H_U$ between two-level systems on adjacent sites of the honeycomb lattice,
we label the three different link directions of the honeycomb lattice by $\alpha \in \{x,y,z\}$, see \cref{fig:model}(b).
On each site $s$, we partition the four two-level systems into two classes $L_{s}^\alpha$ and $R_{s}^\alpha$:
\begin{align}
  \begin{aligned}
    L_{s}^x & = \{1,2\}, \quad & R_{s}^x & = \{3,4\}\,, \\
    L_{s}^y & = \{2,3\}, \quad & R_{s}^y & = \{1,4\}\,, \\
    L_{s}^z & = \{1,3\}, \quad & R_{s}^z & = \{2,4\}\,.
  \end{aligned}
  \label{eq:partitions}
\end{align}
Then, two-level systems $i$ and $j$ on adjacent sites $s$ and $s'$ along an $\alpha$-link of the honeycomb
lattice interact via the blockade interaction $U$,
if and only if $i \in L_{s}^\alpha$ and $j \in R_{s'}^\alpha$ or vice versa. Hence the Hamiltonian $\HU$ takes the form
\begin{align}
  \HU = U \sum_{\langle s,s' \rangle_\alpha} \left[
    \sum_{\substack{i \in L_{s}^\alpha, j \in R_{s'}^\alpha}} \!\!\!\! n_i\, n_j
    \:\: + \!\!\!\!\!\sum_{\substack{i \in R_{s}^\alpha, j \in L_{s'}^\alpha}}    \!\!\!\!
    n_i\, n_j  \right]\,,
  \label{eq:HU}
\end{align}
see \cref{fig:model}(b) for the illustration along an $x$-link. We point out that introducing
two different blockade interaction strengths $V$ and $U$ is not essential for the physics discussed in this work.
However, it is convenient for the numerical simulations, as it allows us to reduce the Hilbert space by taking the
limit $V \to \infty$ and restricting the Hilbert space to zero or a single excitation per site of the honeycomb lattice.

The classical ground state configurations at $\Omega =0$ for $\Delta>0$ and $V\rightarrow \infty$ are given by a
single excitation per site of the honeycomb lattice, and satisfy the blockade interaction $U$ between adjacent sites.
These configurations can be mapped to a $\mathbb{Z}_2$ loop model on the honeycomb lattice by the following rule:
we draw a line connecting two neighboring sites $s$ and $s'$ along an $\alpha$-link if the excitation on site $s$
is in $L_{s}^\alpha$, and the excitation on site $s'$ is in $L_{s'}^\alpha$. For every site, this condition
is satisfied on exactly two links if the excitation is on vertex 1, 2, or 3, and on zero links if it is on vertex 4.
Hence this construction gives a one-to-one mapping between a classical ground state configuration
and loop configurations -- which satisfy the toric code constraint on each site of the honeycomb lattice, i.e., a loop
can never end and never branch.

\subsection{Full symmetry of the blockade structure}
\label{sec:relation}

The Hamiltonian~\eqref{eq:H} exhibits a local symmetry derived from the graph automorphisms of the blockade
structure. The latter are realized by a permutation of the four two-level systems on each site around a plaquette of the
honeycomb lattice, see \cref{fig:model}(c). On each site $s$, we define the permutations $C^{\alpha}_s$ as
double transpositions of the two-level systems:
\begin{align}
  C^{x}_s = (1\,2)(3\,4)\,, \quad
  C^{y}_s = (1\,4)(2\,3)\,, \quad
  C^{z}_s = (1\,3)(2\,4)\,. \nonumber
\end{align}
Note that each permutation $C^{\alpha}_s$ leaves the partition $L_{s}^{\alpha}$ and $R_{s}^{\alpha}$ of
the four two-level systems on site $s$ invariant, while it exchanges $L_s^{\beta}$ and $R_s^{\beta}$
for $\beta \neq \alpha$. These permutations $C_s^{\alpha}$ on the graph lead to a corresponding representation
on the Hilbert space of the two-level systems, which we also denote by $C_s^{\alpha}$. It is then obvious
that the permutations $C_s^{\alpha}$ commute with the terms $\HOm$,
$H_\Delta$, and $H_V$ of the Hamiltonian. On the other hand,
the term $H_U$ is invariant only under the following product of permutations $C_s^{\alpha}$ along
a plaquette of the honeycomb lattice
\begin{align}
  U_p = \prod_{s\in \partial p} C^{\alpha_s}_s
  \label{eq:Op}
\end{align}
with $\alpha_s$ the direction of the external link at site $s$, see \cref{fig:model}(c).
Therefore, the local symmetry $U_p$ commutes with the full Hamiltonian, $[U_p, H] = 0$. Furthermore,
these operators satisfy $U_p^2 = \mathds{1}$ and all $U_p$ mutually commute, i.e., $[U_p, U_{p'}]=0$.
Hence they give rise to a local $\mathbb{Z}_2$ symmetry. It is this local symmetry which maps
the classical ground state configurations onto each other and gives rise to the \emph{full symmetry}
of this blockade structure as defined in Ref.~\cite{Maier2025TopologicalOrder}. From the analysis in
Ref.~\cite{Maier2025TopologicalOrder}, and the general properties of fully symmetric blockade structures,
it then follows that the ground state of the Hamiltonian~\eqref{eq:H} at finite $\Omega\neq 0$ is unique and in
the symmetric sector with $U_p = +1$ for all plaquettes $p$.

The spectrum of the Hamiltonian $H$ splits into different symmetry sectors labeled by the eigenvalues $U_p = \pm 1$
of the plaquette operators. We associate a state with $U_p = - 1$ at a plaquette $p$ with a charge excitation
which is static and does not move under the dynamics of the Hamiltonian. Note that $\prod_p U_p = \mathds{1}$ on the torus (periodic boundaries)
so that the number of charge excitations is always even. The energy difference
between the ground state in the symmetric sector and the lowest energy state in a sector with two charge
excitations at plaquettes $p$ and $p'$ is the charge gap $\dgap{c}$; the latter depends on the distance $d$
between the two charges at $p$ and $p'$.

For technical reasons, it will be convenient to add an additional auxiliary term
\begin{align}
  \HK = K \sum_{p} \bigl( \mathds{1} - U_p \bigr)
  \label{eq:HK}
\end{align}
to the Hamiltonian~\eqref{eq:H}. Since all operators $U_p$ commute with the original Hamiltonian, this auxiliary term does not change the
\emph{eigenstates} of the Hamiltonian; it also leaves the ground state energy invariant.
However, it spreads the spectrum between charge sectors and thereby helps to enforce the symmetry of the ground state in numerical simulations;
in particular, it allows us to target the excitation gaps in the different symmetry sectors.

\subsection{Relation to the toric code blockade structure}

The blockade model presented here is closely related to the toric code
blockade structure introduced in Ref.~\cite{Maier2025TopologicalOrder}. In the following,
we discuss this relation. This discussion is not relevant for the remainder of the paper,
and the reader who is not familiar with the previous work can skip to \cref{sec:effective}.

The above construction of the Hamiltonian \eqref{eq:H} can be motivated by and derived for
$U=V$ from the more complicated blockade Hamiltonian
introduced in Ref.~\cite{Maier2025TopologicalOrder}. Namely by eliminating the two-level systems on
the links of the honeycomb lattice by the following procedure (cf.\ Fig.~6 in Ref.~\cite{Maier2025TopologicalOrder}):
In $H$, vertices on two adjacent sites $s$ and $s'$ exhibit a blockade interaction of strength $U$
if and only if they are \emph{not} vertices of distance 2 in the graph of the original toric code blockade Hamiltonian, i.e.,
they are not directly connected via a link vertex.
Then, the mapping of the classical ground state configurations onto loop configurations on the honeycomb lattice
is in one-to-one correspondence with the mapping of the classical ground state configurations of the original toric
code blockade structure: an edge of our loop configuration is present on an $\alpha$
link if and only if the corresponding link vertex of the original toric code blockade structure would be excited.
Therefore the model introduced here can be viewed as a reduced version of the toric code blockade structure
from Ref.~\cite{Maier2025TopologicalOrder} -- which allows for more efficient numerical simulations due to a
significant reduction of the Hilbert space dimension within one unit cell of the honeycomb lattice.
From this observation, we can immediately conclude that our model is fully symmetric in the sense of
Ref.~\cite{Maier2025TopologicalOrder}, and the proofs therein can be straightforwardly adapted to our model. In particular,
it follows that the ground state of $H$ is unique and always in the symmetric sector with $U_p = +1$ for all plaquettes $p$.
The proof of the existence of topological order in this ground state for weak driving $\Omega$ carries over as well and
it follows that there is a finite gap $\dgap{s}$ for the excitations in the symmetric sector. Furthermore,
the proof of the existence of a charge gap \cite{Fell2026} carries over to the new Hamiltonian $H$.
We emphasize that all these facts are not crucial for the numerical analysis below -- we will verify all of them independently.
However, the analytical backbone verifies and supports the interpretation of the numerical results.

\subsection{Preparation for iDMRG simulations and global symmetries}
\label{sec:effective}

\begin{figure}[tb]
  \centering
  \includegraphics[width=\columnwidth]{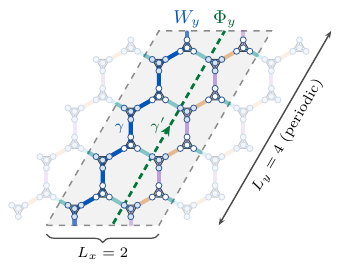}
  \caption{\textit{Geometry for the iDMRG simulations and loop operators.}
    Illustration of the infinite cylinder with the iMPS unit cell (dashed box) of size $L_x = 2$
    and $L_y = 4$ combining $L_x L_y = 8$ unit cells of the honeycomb lattice.
    We have periodic boundary conditions along the $y$-direction. The blue line along the $y$-direction denotes
    the loop operator $W_y$ associated with winding a flux around the cylinder,
    while the green dashed line denotes the Wilson loop operator $\Phi_y$ winding a charge
    around the cylinder. The latter acts with $Q_s^{\alpha}$ on the right site
    of the $\alpha$-link crossed by the loop operator. The operator $W_y$ commutes with the Hamiltonian
    and we select the topological sector with $W_y=1$. By contrast, the Wilson loop $\Phi_y$ exhibits a perimeter law for
    minimally entangled states in the topological phase and vanishes in the trivial phase.
  }
  \label{fig:geometry}
\end{figure}

In this section, we specialize and reformulate the system for the numerical
iDMRG simulations on an infinite cylinder with a finite circumference $L_y$; we
focus on a unit cell of size $L_x = 2$ in $x$-direction, see
\cref{fig:geometry}.

\subsubsection{Hilbert space reduction}

In a first step, we reduce the local Hilbert space on each site by taking the limit $V \to \infty$. This limit removes all states
with two or more excitations on a single site of the honeycomb lattice. Then, the local Hilbert space $\mathcal{H}_s$ reduces to
a five-dimensional Hilbert space with the following basis states: $\ket{0}_s$ is the empty state with all two-level
systems in state $\ket{0}_i$, and $\ket{i}_s$ denotes the state with a single excitation on the $i$-th two-level system
on site $s$.

\subsubsection{Local basis change}

While the above basis is diagonal in the operators $n_{i}$, it is convenient to introduce a different basis
which is diagonal in the operators $C_s^{\alpha}$. Note that the three operators $C_s^{\alpha}$
combined with the identity $\mathds{1}$ form the abelian group $G=\mathbb{Z}_2 \times \mathbb{Z}_2$ (the Klein four-group).
Its irreducible representations are one-dimensional and labeled by $q = (q_x, q_y) \in \mathbb{Z}_2 \times \mathbb{Z}_2$
with the character $\chi_q\left(C_s^{\alpha}\right) = (-1)^{q_\alpha}$; here we defined $q_z = q_x + q_y \mod 2$.
Clearly, the empty state $\ket{0}_s$ is invariant under the action of all $C_s^{\alpha}$ and therefore carries the trivial representation $q=(0,0)$. By contrast, the four states with a single excitation
\begin{align}
  \ket{e_q}_s \equiv \frac{1}{2} \left[\mathds{1}+\sum_{\alpha} \chi^{*}_q\bigl(C_s^{\alpha}\bigr)\, C_s^{\alpha} \right] \ket{1}_s
  \label{eq:irrep-basis}
\end{align}
each carry the representation $q$, i.e., they are eigenstates of the operators $C_s^{\alpha}$ with eigenvalues
$\chi_q\left(C_s^{\alpha}\right) = (-1)^{q_\alpha}$. Thus we can associate with each basis state
a Klein charge $q$ which is probed by the operators $C_{s}^{\alpha}$. With the above choice of $C^{\alpha}_s$,
the new basis states take the explicit form
\begin{eqnarray}
  \ket{e_{0,0}}_s & = & \left(\ket{1}_s+\ket{2}_s+\ket{3}_s+\ket{4}_s\right)/2,  \nonumber\\
  \ket{e_{1,0}}_s & = & \left(\ket{1}_s - \ket{2}_s - \ket{3}_s+\ket{4}_s\right) /2, \nonumber\\
  \ket{e_{0,1}}_s & = & \left(\ket{1}_s+\ket{2}_s-\ket{3}_s-\ket{4}_s\right) /2 , \nonumber\\
  \ket{e_{1,1}}_s & = & \left(\ket{1}_s-\ket{2}_s+\ket{3}_s-\ket{4}_s\right) /2. \nonumber
\end{eqnarray}

In this basis, the local symmetry $U_p$ becomes diagonal while the blockade interaction $\HU$ is off-diagonal.
However, the blockade interaction can conveniently be expressed by introducing the operators $Q_s^{\alpha}$
\begin{align}
  Q_s^{\alpha} & \equiv
  \ket{1}\bra{1}_s+\sum_{\beta} \chi_{m_\alpha}\bigl(C_s^{\beta}\bigr) \, C_{s}^{\beta }\ket{1}\!\bra{1}_s C_{s}^{\beta}
  \label{eq:Qdef}       \\
  & =
  \sum_q  \ket{e_{q+m_\alpha}}\!\bra{e_q}_s\,,
  \qquad
  m_\alpha = \begin{cases}
    (0,1), & \alpha = x, \\
    (1,0), & \alpha = y, \\
    (1,1), & \alpha = z.
  \end{cases}  \nonumber
\end{align}
Then, the Hamiltonian~\eqref{eq:H} can be written in the limit $V\rightarrow \infty$
and with the auxiliary term $\HK$ in \cref{eq:HK} as (see \cref{app:irrep} for details)
\begin{align}
  H & =
  \Delta \sum_s \bigl( \mathds{1} - n_s \bigr)
  + \frac{U}{2} \sum_{\langle s, s' \rangle_\alpha}
  \Bigl( n_s\, n_{s'} - Q^{\alpha}_s\, Q^{\alpha}_{s'} \Bigr)
  \label{eq:H-irrep}                                                                              \\
  & \phantom{=\,} + 2\,\Om \sum_s \bigl( \ket{e_{00}}\!\bra{0}_s+\ket{0}\!\bra{e_{00}}_s \bigr)
  + K \sum_p \bigl( \mathds{1} - U_p \bigr)\,. \nonumber
\end{align}
Note that the blockade interaction exchanges the Klein charge $m_\alpha$ of the irreducible representations
between two adjacent sites along the $\alpha$-link. On the other hand, the local plaquette symmetry
$U_p$ is diagonal in this basis and the Rabi term $\HOm$ only couples the two states carrying
the trivial representation, i.e., the empty state $\ket{0}_s$ and the state $\ket{e_{00}}_s$.

Note that the operators $Q_s^{\alpha}$ and $C_s^{\alpha}$ satisfy a simple algebraic structure.
First, they obey the constraints $Q_s^{x} Q_s^{y} Q_s^{z} =n_s$ and $C_s^{x} C_s^{y} C_s^{z} = \mathds{1}$;
furthermore, they obey the commutation relations
\begin{align}
  C_s^{\alpha} Q_s^{\beta}  + (-1)^{\delta_{\alpha,\beta}} Q_s^{\beta} C_s^{\alpha} = 0\,,
  \label{eq:commutation}
\end{align}
i.e., $C_s^{\alpha}$ commutes with $Q_s^{\alpha}$ but anti-commutes with $Q_s^{\beta}$
for $\beta \neq \alpha$. It is exactly this commutation relation that shows that the Hamiltonian $H$
is invariant under the local symmetry $U_p$ in~\cref{eq:Op}.

\subsubsection{Flux loop operator}

In addition to the local symmetries $U_p$, one can also define a global symmetry associated with the
non-trivial loops winding around the cylinder. Of special interest is the operator $W_y$ associated with a
loop $\gamma$ on the honeycomb lattice winding around the circumference of the cylinder, see the blue line in \cref{fig:geometry}.
This operator is defined as the product
\begin{align}
  W_y \equiv \prod_{s \in \gamma} C_{s}^{y}
  \quad\text{with}\quad
  [W_y, H] = 0\,.
  \label{eq:Wy}
\end{align}
From the full symmetry of the Hamiltonian $H$ it follows that the ground state at finite $\Omega \neq 0$
is an eigenstate of $W_y$ in the sector with $W_y = +1$. For the iDMRG simulations we therefore fix an initial state with $W_y=+1$
and also observe that the final converged ground state is in the sector with $W_y = +1$.

\subsubsection{Wilson loop operator}

The fixed point of the toric code also is an eigenstate of the Wilson loop operator $\Phi_y$
associated with a loop $\gamma'$ winding along the dual lattice around the cylinder,
see the green dashed line in \cref{fig:geometry}. It is defined as the product of operators $Q_s^{\alpha}$
\begin{align}
  \Phi_y \equiv \prod_{s \in \gamma'} Q_{s}^{\alpha_s}\,.
  \label{eq:PhiA}
\end{align}
Note that the loop $\gamma'$ on the dual lattice crosses the links of the honeycomb lattice and is a directed loop.
Then, $s \in \gamma'$ means that $s$ is the site on the link crossed by $\gamma'$, which is on the right side of the loop $\gamma'$.
In addition, $\alpha_s$ denotes the label of the crossed link. Therefore the particular operator $\Phi_y$ depicted in
\cref{fig:geometry} is a product of four operators $Q_s^{y}$ on the sites touched by the green line.
Note, that changing the direction of the loop gives rise to a different operator -- however, the two operators are connected
by inversion symmetry of the honeycomb lattice, and therefore have the same expectation value for all ground states observed here.

The Wilson loop operator $\Phi_y$ is \emph{not} a symmetry of the Hamiltonian $H$ (except at the fixed point $\Omega =0$). However,
it is known that for iDMRG simulations on an infinite cylinder, the \emph{minimally entangled states} of a topological phase are
characterized by a finite expectation value $|\langle \Phi_y \rangle|>0$ \cite{Zhang2012MES}. Importantly, minimally entangled
states are required for the extraction of the topological entanglement entropy on an infinite cylinder. Furthermore,
iDMRG is known to prefer these states \cite{Jiang2012Identifying,Cincio2013Characterizing}
so that the converged ground state of the iDMRG simulations is expected to be a minimally entangled state.
Hence we use the expectation value of the Wilson loop operator
$\Phi_y$ as an indicator of the topological order and will show that the vanishing of $|\langle \Phi_y \rangle|$ coincides
with the phase transition from the topologically ordered phase into the trivial phase. Note that the \emph{sign}
of $\langle \Phi_y \rangle$ is spontaneously selected during the iDMRG simulations -- but once a sign is selected, it remains fixed
for the entire run, i.e., starting with an initial state with a fixed value of $\langle \Phi_y \rangle$,
the iDMRG simulation converges to a state with the same sign of $\langle \Phi_y \rangle$.
Consequently, the sign of $\langle \Phi_y \rangle$ and the sign of $W_y$ fully select the
topological sector in the ground state manifold.

\subsubsection{Global $\mathbb{Z}_2 \times \mathbb{Z}_2$ symmetry}

The local symmetries of the Hamiltonian can be combined into global symmetries, which
allow one to reduce the computational cost of the iDMRG simulations. Here, we
are interested in the global $\mathbb{Z}_2 \times \mathbb{Z}_2$ symmetry
generated by the operators
\begin{align}
  \mathcal{K}^{\alpha} \equiv \prod_s C_{s}^{\alpha}\,.
  \label{eq:vglob}
\end{align}
It corresponds to the conservation of the global Klein charge $(k_x,k_y)$, the
irreducible representations of this global symmetry group.
In the irreducible basis $\{\ket{0}_s,\ket{e_q}_s\}$ of the local Hilbert spaces, these operators are
diagonal, and it is straightforward to impose the conservation of $(k_x,k_y)$ during
iDMRG simulations.

However, there is one important caveat. Fixing the conservation of the global charge $k_x$
forces the ground state for \emph{odd} $L_y$ into a superposition of topological sectors with $\langle \Phi_y \rangle=0$, i.e.,
the converged ground state from iDMRG is no longer a minimally entangled state -- which can lead to an
incorrect extraction of the topological entanglement entropy.
One can understand this as follows: For even $L_y$, the global operator $\mathcal{K}^x$ can be
written as a product of plaquette operators $U_p$. For odd $L_y$, however, $\mathcal{K}^x$ necessarily involves the operator
$W_x =\prod_{s\in \gamma} C_s^{z}$ for a path $\gamma$ along the cylinder; setting $k_x=0$ then fixes
the topological sector to $W_x=1$.
Since the operator $W_x$ anticommutes with the Wilson loop operator $\Phi_y$,
this forces the ground state into a superposition of two topological sectors
such that $\langle \Phi_y \rangle=0$; in particular, the converged ground
state is no longer a minimally entangled state \cite{Zhang2012MES}.
As a consequence, we only fix $k_y=0$ in our simulations, which allows
the iDMRG to converge into minimally entangled ground states.

\begin{figure*}[tb]
  \centering
  \includegraphics{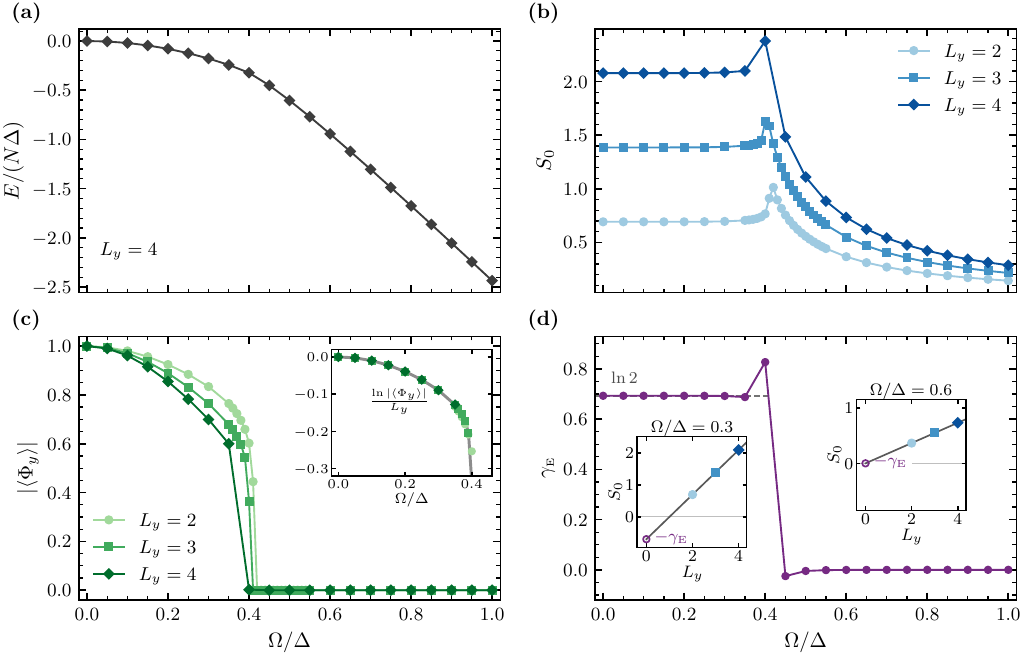}
  \caption{
    \textit{Topological phase transition.}
    (a) Ground-state energy $E/N$ per unit cell of the honeycomb lattice for the blockade Hamiltonian in the limit $V\rightarrow \infty$
    and for parameters $U=2 \Delta$ using iDMRG on a cylinder of size $L_x=2$ and $L_y=4$. The ground state energy behaves smoothly
    across the phase transition and no signature of hysteresis is observed.
    (b) Entanglement entropy $S_0$ across the cylinder for different circumferences $L_y = 2, 3, 4$; one finds a clear
    signature of a phase transition at a critical value $\Omc/\Delta \approx 0.40$. The entanglement entropy for $\Omega \ll  \Omega_c$
    is close to $S_0 = (L_y - 1)\ln 2$, which is the entanglement entropy of the toric code state at the fixed point.
    (c)  Expectation value of the Wilson loop operator $\Phi_y$, which is finite for $\Omega \leq \Omega_c$ and drops to zero at the phase transition.
    The inset shows that it exhibits a perimeter law scaling $|\langle \Phi_y\rangle| \sim e^{-L_y/\xi(\Om)}$ for increasing circumference
    $L_y$ of the cylinder, as expected for a topological phase.
    (d)  Topological entanglement entropy $\tee$ extracted from the linear fit $S_0 = a L_y - \tee$ for the entanglement entropy
    through $L_y = 2, 3, 4$. The insets show two examples of the linear fit at $\Om/\Delta = 0.3$ and $\Om/\Delta = 0.6$.
    The extracted topological entanglement entropy is $\tee \approx \ln 2$ for $\Omega< \Omega_c$ and drops to zero above the critical point $\Omc$.
    These observations demonstrate a phase transition from a topological phase with toric code order to a trivial phase for $\Omega > \Omc$.
  }
  \label{fig:transition}
\end{figure*}

\section{Numerical results}
\label{sec:phasediagram}

\subsection{Ground state properties and topological phase transition}

In the following, we study the phase diagram and the transition from the topological phase to the trivial phase for
our model using extensive iDMRG simulations. We focus on the reduced model in the regime $V\rightarrow \infty$
with five states per site. Then, the Hamiltonian is given by \cref{eq:H-irrep} and we focus
on the parameters $U = K =2 \Delta$ for varying Rabi drive $\Om$. We use the open source software package \texttt{TeNPy}~\cite{Hauschild2024TeNPy}
to perform the simulations.
In a first step, we derive the ground state properties using iDMRG on an infinite cylinder with a unit cell for the
infinite-size matrix product state (iMPS) of size $L_x=2$ and circumference of $L_y = 2,3,4$, see \cref{fig:geometry}.
We increase the bond dimension $\chi$ up to $\chi=1536$ until convergence of the ground state energy and the entanglement entropy is reached
with a truncation error below $10^{-8}$.
For $L_y=2$, the bond dimension $\chi=256$ ($\chi=512$ close to the transition) is sufficient, while for $L_y=4$ up to
$\chi=1536$ is necessary. We implement the global conservation of the charge $k_y=0$ but, as discussed above, do
not require the conservation of $k_x$. The auxiliary term $\HK$ with $K=U$ is included to obtain
fast convergence towards the symmetric sector; the stability of the ground state is tested for different values of $K$,
in particular for $K=0$. From the analytical analysis, we know that the ground state is in the symmetric sector with $\langle \HK \rangle = 0$.
Hence we can use the expectation value $\langle \HK \rangle$ as a self-consistency check for the convergence, and
as an estimator on the residual error of the ground state energy. In addition, the initial state selects the sector $W_y=+1$
which is kept during the iDMRG simulations; this expectation value is also checked for consistency for all results shown here.

The ground state energy per unit cell of the honeycomb lattice and the entanglement entropy
along a cut of the cylinder in $y$-direction are shown in \cref{fig:transition}(a,b)
as a function of the Rabi drive $\Om$. The entanglement entropy shows a clear signature of a phase
transition at a critical value $\Omc/\Delta \approx 0.40$. We performed scans across the transition in
both directions, where the initial state for each $\Omega$ was chosen as the converged ground state of
the previous parameter point, and found a smooth behavior in the ground state energy without any signs
of hysteresis or metastability, see \cref{fig:transition}(a). These signatures are therefore consistent with a continuous
second-order phase transition.

\begin{figure*}[tb]
  \centering
  \includegraphics{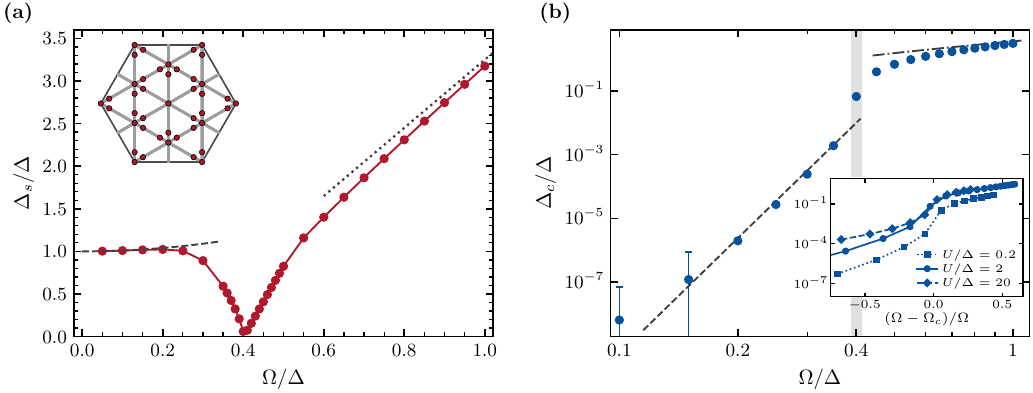}
  \caption{
    \textit{Excitation gaps.}
    (a) The gap $\Delta_s$ of excitations in the symmetric sector with $U_p=+1$ for
    all plaquettes for a cylinder of circumference $L_y=3$. The gap is derived via iDMRG using
    the plane-wave ansatz implemented in \texttt{TeNPy}. We probe the momenta $p_x =0, \pi/2, 2\pi/3$ along
    the $x$-direction.
    In combination with the finite-size quantization along the $y$-direction and the point group symmetry of the
    honeycomb lattice, this leads to the explored momentum points in the first Brillouin zone shown in the inset.
    The gap $\Delta_s$ is the minimum over all excitation energies at the explored momenta. The dashed and dotted
    lines show the analytical predictions for the gap $\Delta_s$ in the limit of small and large $\Omega$, respectively.
    The gap $\Delta_s$ is a lower bound for the gap of flux anyons in the topological phase; the closing
    of the gap is then consistent with the condensation of flux anyons at the transition.
    (b) The charge gap $\Delta_c$ is determined as the energy difference of the ground state to the lowest energy state
    with two charge excitations ($U_p=-1$) at two plaquettes separated by a distance $d=2$ along the
    $y$-direction. The dashed line shows the analytical prediction for the charge gap in the limit of small $\Omega$,
    while the dash-dotted line shows the analytical prediction for large $\Omega$. The grey bar indicates the transition
    $\Omc/\Delta \approx 0.40$, while for the error bars we use the residual energy per unit cell $|\langle H_K\rangle|/N$;
    this quantity vanishes for the exact lowest energy state in a specific symmetry sector, and therefore provides
    a lower bound on the uncertainty in energy per unit cell.
    Remarkably, the charge gap remains finite across the transition. The inset shows a comparison of the charge gap for different
    strengths of the interaction $U=\{0.2,2,20\}\Delta$ as a function of the rescaled distance from the
    phase transition.
  }
  \label{fig:gaps}
\end{figure*}

The behavior of the topological entanglement entropy $\tee$~\cite{Kitaev2006TEE,Levin2006Detecting} and the Wilson loop operator $\Phi_y$
is shown in \cref{fig:transition}(c,d); we find for $\Omega < \Omega_c$ a topological phase consistent with toric code topological order,
as predicted from the analytical analysis in the limit $\Omega \to 0$. The Wilson loop operator $\Phi_y$ is expected to exhibit a perimeter
law in the topological phase \cite{HastingsWen2005}. (Note that here we have a local $\mathbb{Z}_2$ symmetry, as in pure lattice gauge
theories without charge fluctuations \cite{Wegner1971,FradkinSusskind1978}.) Therefore the expectation value $|\langle \Phi_y \rangle|$
of the Wilson loop operator vanishes for $L_y \to \infty$. Such a behavior is confirmed by a finite size scaling analysis in the inset
of \cref{fig:transition}(c) with $\ln |\langle \Phi_y \rangle|/L_y$ collapsing onto a single curve for all circumferences $L_y$.
However, at finite $L_y$, $|\langle \Phi_y \rangle|$ serves as a good order parameter for the topological phase transition~\cite{Trebst2007Breakdown}.
By contrast, for $\Omega > \Omega_c$ we find a trivial phase with vanishing topological entanglement entropy and vanishing Wilson loop
operator $\Phi_y$. Such a behavior is expected in the trivial phase where the Wilson loop operator $\Phi_y$ follows
an area law~\cite{Gregor2011Diagnosing}. Note that on a cylinder the
loop $\gamma'$ can be understood as the limit of a closed loop with the area of the half-cylinder. The behavior of the trivial phase can
also be understood in the limit $\Omega \rightarrow \infty$, where the ground state is a product of
states $(\ket{0}_s - \ket{e_{00}}_s)/\sqrt{2}$ on each site.
In summary, we find strong numerical evidence for a continuous topological phase transition from a topological phase with toric
code order to a trivial phase at a critical value $\Omc/\Delta \approx 0.40$, in analogy to numerical studies of the toric code in a
magnetic field \cite{Trebst2007Breakdown,Tupitsyn2010Multicritical,Dusuel2011Robustness,Vidal2009LowEnergy,Vidal2009SelfDuality,Wu2012ToricCodeField}.

\subsection{Excitations in the symmetric sector}
\label{sec:excitations}

To further characterize the properties of the ground state across the phase transition,
we study the low energy excitations above the ground state manifold of the model.
First, we focus on the excitation gap $\Delta_s$ within the symmetric sector with $U_p=+1$
for all plaquettes. At the fixed point $\Omega=0$ and in the parameter regime with $U=2 \Delta$,
the first excitation is determined by a single site in the state $\ket{0}_s$ with excitation energy $\Delta$.
Applying lowest order perturbation theory based on a Schrieffer--Wolff transformation~\cite{Bravyi2011}, we find the prediction for
the excitation gap $\Delta_s$ in the limit of small Rabi drive $\Om \ll \Delta$ (see~\cref{app:pt} for details)
\begin{align}
  \Delta_s = \Delta + \frac{\Om^2}{\Delta}
  \,.
  \label{eq:gap-sym-small}
\end{align}
On the other hand, in the polarized phase with $\Om \gg \Delta$, the lowest excitation in the symmetric sector is a single
site in the state $(\ket{0}_s + \ket{e_{00}}_s)/\sqrt{2}$ with excitation energy $4 \Om$.
Using perturbation theory, the leading behavior of the gap $\Delta_s$ for $\Omega \gg \Delta$
is given by (\cref{app:pt})
\begin{align}
  \Delta_s = 4\Om - \frac{3U}{8}
  \,.
  \label{eq:gap-sym-large}
\end{align}
So in both phases we expect a finite excitation gap above the ground state manifold. While in the trivial phase the ground state is unique,
in the topological phase the ground state manifold is nearly degenerate. However, fixing the $W_y=+1$ sector, together with
the condition of a minimally entangled state, fixes the topological sector. Then it is possible to apply the plane-wave excitation
ansatz~\cite{Haegeman2012Variational,Vanderstraeten2019TangentSpace} to determine the lowest excitation gap $\Delta_s$
within the symmetric sector. Such a plane-wave excitation method is provided by the \texttt{TeNPy} package \cite{Hauschild2024TeNPy},
with one caveat: the implementation relies on a uniform iMPS. Therefore, we perform re-optimization using variational uniform matrix
product state (VUMPS)~\cite{ZaunerStauber2018VUMPS}
with the iMPS obtained from iDMRG as an initial state in order to obtain a uniform iMPS.

The numerical results for the excitation gap $\Delta_s$ are shown in \cref{fig:gaps}(a) for a cylinder with
circumference $L_y=3$ and unit cell size $L_x=2$. The gap $\Delta_s$ is determined as the minimum over the sampled points
in the Brillouin zone, which are shown as inset in \cref{fig:gaps}(a); they are determined by the sampled
momenta $p_x =0, \pi/2, 2\pi/3$ along the $x$-direction of the cylinder and the three quantized
momenta across the periodic circumference of the cylinder. The analysis shows a clear signature of a closing of the gap $\Delta_s$
at the transition, which is consistent with a continuous topological phase transition. Furthermore, flux excitations are
excitations in the symmetric sector and $\Delta_s$ provides a lower bound for the energy of these excitations.
The vanishing of $\Delta_s$ at the phase transition is therefore consistent with a condensation
of flux excitations at the transition into the trivial phase \cite{CastelnovoChamon2008,BaisSlingerland2009,Burnell2018AnyonCondensation}.

\begin{figure}[tb]
  \centering
  \includegraphics[width=\columnwidth]{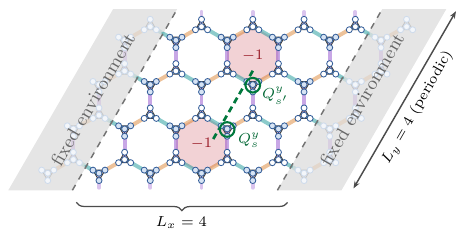}
  \caption{\textit{Setting to probe the charge gap.}
    We select a finite section of length $L_x=4$ of the infinite cylinder with circumference $L_y=4$
    and construct the MPS from the iMPS ground state. The boundary
    conditions are fixed by the left and right environments of the infinite cylinder
    from the iMPS ground state. Then we perform two DMRG simulations on this finite section
    with fixed boundary conditions: one for the ground state in the symmetric sector,
    and one for the lowest energy state with a pair of charges at two plaquettes $p'$ and $p''$,
    separated by a distance $d=2$ along the $y$-direction. To create the two charges, we apply a ribbon operator
    $R= Q^{y}_{s}Q^{y}_{s'}$ on two adjacent sites along the $y$-direction. This operator
    generates two charges at its endpoints. In addition, we modify the auxiliary term $H_K$ in the Hamiltonian
    to select the sector with $U_{p'}=U_{p''}=-1$ (the two red plaquettes).
    The two DMRG runs to determine the lowest energy state in these two symmetry sectors are performed with
    identical settings. Then the charge gap $\Delta_c$ is determined as the energy difference between these two states.
  }
  \label{fig:protocol}
\end{figure}

\subsection{Charge gap}

We can also study the behavior of the charge gap $\Delta_c$ across the phase transition.
The charge gap is defined as the energy difference between the ground state in the symmetric sector
and the lowest energy state with a pair of charges at $p'$, $p''$, i.e., the symmetry sector with
$U_{p'}=U_{p''}=-1$, while $U_p=+1$ for all other plaquettes. In the topological phase, the charge gap
vanishes at the fixed point $\Omega=0$, but a term $1-U_p$ is generated in perturbation theory using
a Schrieffer--Wolff transformation in 12th order. This predicts a charge gap for $U=2 \Delta$ (\cref{app:pt})
\begin{align}
  \Delta_c = C_{12}\, \frac{\Om^{12}}{\Delta^{11}}
  \quad\text{with}\quad
  C_{12} = \frac{1018657}{1728} \approx 589.5\,.
  \label{eq:gap-charge-pt}
\end{align}
On the other hand, in the trivial phase with $\Om \gg \Delta$, the charges are confined with
an energy linearly increasing with the distance $d$ between the two charges. In leading order, perturbation theory
predicts the charge gap at distance $d$
\begin{align}
  \Delta_c(d) = \sigma d
  \quad\text{with}\quad
  \sigma = 2 \Om\,,
  \label{eq:string-tension-pol}
\end{align}
where $\sigma$ denotes the string tension. As we probe the charge gap numerically at distance
$d=2$, we can also evaluate the charge gap for this separation within first order perturbation theory
as $\Delta_c = 4 \Om - \Delta/2$ (\cref{app:pt}).

To probe the charge gap numerically with DMRG, we apply the following protocol, illustrated in \cref{fig:protocol}.
We first construct an MPS for the section of a cylinder with length $L_x=4$ and circumference $L_y=4$ from the iMPS ground state
obtained by the iDMRG simulations on the infinite cylinder, with the boundary conditions fixed by the left and right environments
of the infinite cylinder. From this MPS state,
we construct a new MPS state, with a pair of charges at two plaquettes $p'$ and $p''$ separated by a distance $d=2$, by applying a flux
ribbon operator $R= Q^{y}_{s}Q^{y}_{s'}$ on two adjacent sites along the $y$-direction, see \cref{fig:protocol}. Then we run DMRG
simulations on the finite section with these two initial states and fixed boundary conditions.
Note that for the simulation with the charge pair, the auxiliary term $H_K$
is modified to enforce the sector with a pair of charges at $p'$ and $p''$, i.e., $H_K \mapsto H_K + 2 K \left(U_{p'}+U_{p''}\right)$.
The two runs yield the two lowest energy states in the symmetric sector and the sector with a pair of charges at $p'$ and $p''$, respectively. Both states are computed for the same boundary conditions determined by the iDMRG ground state to avoid boundary effects,
and evaluated with the same bond dimensions up to $\chi=768$ to achieve convergence on the scale of the expected charge gap.
The energy difference between these two states then corresponds to the charge gap $\Delta_c$.

The results for the charge gap $\Delta_c$ using this protocol are shown in \cref{fig:gaps}(b).
As an estimator for the error we use the sum of the expectation values of the plaquette
term $|\langle H_K \rangle|/N$ in both DMRG runs. For a correct eigenstate
in the respective symmetry sector, this expectation value vanishes. Therefore, $|\langle H_K \rangle|/N$
is a lower bound on the error in energy per unit cell within our numerical simulations, and consequently, the gap cannot
be determined with a precision higher than this error.
In addition, the analytical predictions for the charge gap in the topological phase [\cref{eq:gap-charge-pt}] and in the trivial phase
are also shown; we find good agreement with the numerical results. The inset of \cref{fig:gaps}(b) shows a comparison of the charge gap $\Delta_c$ at different
interaction strengths. We emphasize that it is important to study the charge gap on a
cylinder with circumference $L_y\geq4$ to obtain the correct behavior predicted by perturbation theory -- for smaller circumferences
the result is affected by non-trivial loops around the cylinder that appear in the same order of perturbation theory for $L_y=3$, and dominate
the charge gap for $L_y=2$; we have verified these predictions with DMRG simulations for smaller cylinders.
The most important observation is that the
charge gap $\Delta_c$ remains finite across the phase transition.

\subsection{Phase diagram in the \texorpdfstring{$\Om$--$U$}{Omega--U} plane}

\begin{figure}[tb]
  \centering
  \includegraphics{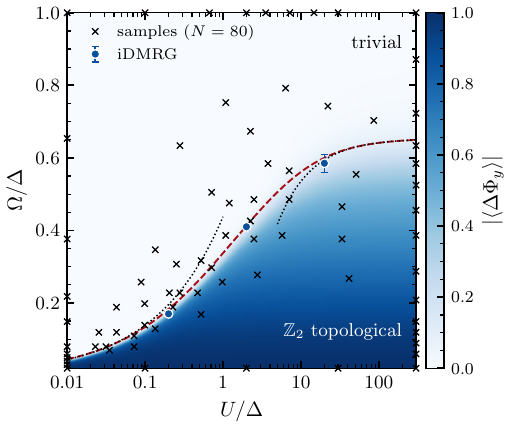}
  \caption{\textit{Full phase diagram.}
    The full phase diagram computed with a reduced basis method using DMRG on a cylinder
    with open boundaries, circumference $L_y=2$, length $L_x=6$ and for
    bond dimension $\chi=256$. The transition of the topological phase to the trivial
    phase is indicated by the observable $|\langle \Delta \Phi_y \rangle|$ and the red dashed line
    indicates the equipotential line $|\langle \Delta \Phi_y \rangle|=0.1$. The blue bullets
    derive from the independent analysis via iDMRG for $U=\{0.2,2,20\}\Delta$ and $L_y=4$
    that was used to derive the behavior of the charge gap for different values of $U$ in \cref{fig:gaps}(b).
    The crosses indicate the samples where the ground state was computed using DMRG; these samples
    are used to iteratively construct the reduced basis sets. After an initial step with a few samples on the boundary, the
    subsequent samples are selected by a greedy algorithm based on the residual error of the reduced basis set. The
    total number of samples was set to $N=80$, at which the phase boundary is already well converged.
    The two dotted lines show the expected behavior of the phase boundary
    in the two asymptotic limits of small and large $U$: $\Omc = \sqrt{g_c U \Delta}$
    with a fitted parameter $g_c = 0.1915(4)$ for small $U$,
    and $\Omc = \Om_\infty (1 - c\,\Delta/U)$ with fitted parameters $\Om_\infty = 0.6533(1)\,\Delta$, $c = 1.81$
    for large $U$.
  }
  \label{fig:phasediagram}
\end{figure}

Finally, we can study the behavior of the critical Rabi drive $\Omc$ as a function of the blockade strength $U$
and derive the full phase diagram in the $\Om$--$U$ plane. For the efficient evaluation of the phase diagram,
we employ a reduced basis method \cite{brehmer2023rb}: the ground state of the Hamiltonian is computed with
DMRG for a set of training points (``samples'') that are used to construct a reduced basis set. The phase diagram in the full plane
is then derived by evaluating the Hamiltonian projected on the Hilbert space spanned by this reduced basis set.
The optimal set of sample points is determined recursively by a greedy algorithm based on the residual
error of the reduced basis set. For each sample, the DMRG ground state is computed on a cylinder of circumference $L_y=2$ and length $L_x=6$
with bond dimension $\chi=256$. Note that the reduced basis method requires DMRG on a finite cylinder with properly normalized states.
As an order parameter, we evaluate the correlator of two Wilson loop operators $\Phi_y(x)$ at position $x=2$ and $x=5$; here $x$ denotes
the position along the cylinder. We use the notation $\Delta \Phi_y = \Phi_y(2) \Phi_y(5)$ for this observable. We use this correlator
as the DMRG on a finite cylinder can converge towards a ground state in a superposition of topological sectors, and therefore
$|\langle\Phi_y \rangle|$ is not a reliable indicator of the phase transition for DMRG on a finite cylinder.
The qualitative phase diagram is shown in \cref{fig:phasediagram},
where the red dashed line indicates the phase boundary and is given by the equipotential line $|\langle \Delta \Phi_y \rangle| = 0.1$;
note that in the trivial phase $|\langle \Delta \Phi_y \rangle|$ is not exactly zero due to the finite size of the cylinder.
The predicted phase boundary is in good agreement with independent iDMRG simulations for fixed $U=\{0.2,2,20\}\Delta$ (blue bullets).
We can also understand the qualitative behavior of the phase boundary in the two asymptotic limits for small and large $U$.
For $U\rightarrow \infty$, the critical Rabi drive $\Omc$ saturates at a finite value $\Om_\infty$ with an expected $1/U$ correction, i.e.,
$\Omc = \Om_\infty (1 - c\,\Delta/U)$. On the other hand, for small $U$, we can adiabatically eliminate the empty state on each
site, leading to an effective hopping of excitations with strength $\Om^2/\Delta$ on each site.
This leads to a behavior $\Omega_c =\sqrt{g_c U \Delta}$. Both expected behaviors are in excellent agreement
with the numerical results, see the black dotted lines in \cref{fig:phasediagram};
we find the fit parameters $g_c = 0.1915(4)$, $\Om_\infty = 0.6533(1)\,\Delta$, and $c = 1.81$.

\section{Summary and Conclusion}
\label{sec:discussion}

In summary, we have presented a detailed numerical study, based on DMRG, of a topological phase transition for a blockade Hamiltonian.
The model is a simplified version of the toric code blockade Hamiltonian introduced in Ref.~\cite{Maier2025TopologicalOrder}
that still exhibits a local $\mathbb{Z}_2$ symmetry, but where the two-level systems on the links have been eliminated.
Due to the close relation between the two models,
the mathematically rigorous proofs on the existence of topological order for weak Rabi driving carry over
to the simplified model studied here. The numerical analysis confirms this prediction and shows that the model exhibits a robust
topological phase with toric code order up to a critical Rabi drive $\Omc$ of order $\Delta$.
Furthermore, the analysis demonstrates that the topological
phase is gapped, with a gap for both anyonic excitations (charges and fluxes). The charge gap remains finite across
the phase transition and the charges become confined in the trivial phase, whereas the numerical analysis is consistent with
the condensation of flux excitations and a vanishing flux gap at the phase transition. These results are consistent
with analytical predictions based on the general anyon-condensation framework
for such topological phase transitions \cite{BaisSlingerland2009, Burnell2018AnyonCondensation}.

The observation of a finite charge gap across the phase transition is of particular interest
for potential experimental realizations: one can expect that the topological phase remains robust to small
perturbations that break the local $\mathbb{Z}_2$ symmetry, as long as the perturbations are smaller
than the charge gap. Such symmetry-breaking perturbations are unavoidable in any experimental realization.
Important open questions are whether the charge gap can be made larger by modifying the blockade Hamiltonian,
and whether there are experimentally realistic schemes for implementing such Hamiltonians.

Finally, we would like to point out, that the phase diagram of the blockade Hamiltonian studied here is also accessible
via quantum Monte Carlo simulations, as the blockade graph does not give rise to frustration. Such an analysis would
allow one to gain further insights into the thermal properties of the quantum many-body systems, as well as extend the system
to larger transversal system sizes. We have tested such quantum Monte Carlo simulations, and first preliminary results
are fully consistent with the zero temperature results based on iDMRG presented in this work.

\begin{acknowledgments}
  We thank Benjamin Stamm for discussions on the reduced basis method.
  L.P. acknowledges financial support from the Deutsche Forschungsgemeinschaft (DFG),
  project Nr 530111096. The project/research is part of the Munich Quantum Valley,
  which is supported by the Bavarian state government with funds from the Hightech Agenda Bayern Plus.

  \textit{Note on AI usage.} Supervised AI agents performed all analytical calculations, implemented and conducted all numerical simulations,
  and created all figures in this manuscript. The following LLMs have been used: Claude Fable (version 5.0 and 5.1) and Claude Opus (version 4.8) by Anthropic.
  The authors tested and verified all results independently, reviewed the code, performed consistency checks, and therefore take full responsibility for the content of this manuscript.
  The text of the manuscript was written by the authors and only proofread by the AI agent.
\end{acknowledgments}

\textit{Data availability statement.} The data that support the findings of this article are openly available \cite{Moegerle2026ReplicationData}.

\bibliography{bibliography}


\appendix

\crefalias{section}{appendix}

\section{Blockade interaction in the irreducible representation basis}
\label{app:irrep}

In the following, we express the blockade interaction $\HU$
in \cref{eq:HU} in terms of the operators $Q_s^{\alpha}$ introduced in \cref{eq:Qdef}. This leads to
the simple form of the Hamiltonian in \cref{eq:H-irrep}.

First, the operator $Q_s^{\alpha}$ is diagonal in the basis $\ket{i}_s$,
$Q_s^{\alpha} = \sum_{i} q^{\alpha}_i \ket{i}\!\bra{i}_s$ with eigenvalues $q^{\alpha}_i = \chi_{m_\alpha}(g_i) = \pm 1$,
where $g_i$ denotes the unique permutation that maps the vertex $1$ onto the vertex $i$, i.e.,
$(g_1, g_2, g_3, g_4) = (\mathds{1}, C_s^{x}, C_s^{z}, C_s^{y})$. In addition, $m_\alpha$ is uniquely defined by the condition
$\chi_{m_\alpha}(C_s^{\alpha}) = +1$. Therefore, we obtain $q^{\alpha}_i = +1$ for the two vertices $i \in \{1, C_s^{\alpha}(1)\}$,
which form one class of the partition for the link $\alpha$, while $q^{\alpha}_i = -1$ for the two vertices in the other class.
As a consequence, we find that two excitations at $i \in s$ and $j \in s'$ on adjacent sites along an $\alpha$
link are in different classes if and only if $q^{\alpha}_i q^{\alpha}_j = -1$. Then, the projector
onto pairs of vertices on adjacent sites in different classes takes the form
\begin{align}
  P^{\alpha}_{s,s'} = \sum_{i\in s, j\in s'} \frac{1 - q^{\alpha}_i q^{\alpha}_j}{2} \, n_i n_j
  =\frac{1}{2} \left( n_s n_{s'} - Q^{\alpha}_s Q^{\alpha}_{s'}  \right) \nonumber
\end{align}
with $n_s = \sum_{i \in s} n_i$ and $Q_s^{\alpha} = \sum_{i \in s} q^{\alpha}_i\, n_i$.
However, the blockade interaction $\HU$ in \cref{eq:HU}
is exactly the sum of all these projectors over adjacent sites $s$ and $s'$, i.e.,
\begin{align}
  \HU= U \sum_{\langle s, s' \rangle_\alpha} P^{\alpha}_{s,s'}\,.
\end{align}
%

\section{Perturbation theory}
\label{app:pt}

In this appendix, we provide the detailed calculations for the perturbative expansions
for small $\Omega \ll \Delta$ and large $\Omega \gg \Delta$.
We focus on parameters $U = 2\Delta$ and $V\rightarrow \infty$
throughout this appendix. The perturbative expansion is always performed in a fixed symmetry sector.

\subsection{Weak Rabi coupling $\Om \ll \Delta$}

We start by deriving the gap $\Delta_s$ in the symmetric sector for small $\Om\ll \Delta$ using perturbation theory
in the form of a Schrieffer--Wolff transformation up to second order in $H_\Om$. Since the gap is defined as the energy difference
between the ground state and the first excited state in the symmetric sector, we need to evaluate the energy
corrections for both states. It is important to note that the first and third order corrections
vanish as $H_\Om$ always changes the total number of excitations. Then, applying standard second order perturbation theory
based on a Schrieffer--Wolff transformation, we find the second order energy correction for the ground state $\ket{\psi_0}$
with energy $E_0=0$ as
\begin{align}
  \frac{E(\Omega)}{N}
  =  - \frac{2 \Omega^2}{\Delta}
\end{align}
since the second order term virtually excites the empty state $\ket{0}_s$ with energy cost $\Delta$ on a site $s$.
Here, $N$ is the number of unit cells in the honeycomb lattice. In the parameter regime $U=2 \Delta$, the lowest excitation
is characterized by a single site $s$ in the empty state $\ket{0}_s$. Then, we obtain
again the same terms as for the ground state for sites not involving $s$, however, at site $s$ we have four different virtual excitations:
one with energy cost $\Delta$, and three with energy cost $\Delta-2 U$ due to the blockade interaction.
Therefore, we obtain the gap
\begin{align}
  \Delta_s
  = \Delta + \frac{2 \Omega^2}{\Delta} +\frac{3 \Omega^2}{\Delta-2 U}
  = \Delta + \frac{\Omega^2}{\Delta}\,.
\end{align}
Note that a potential hopping term of the excitation in second order cancels
due to the destructive interference of the two possible paths; the dispersion
relation of these excitations is only generated in fourth order.

Next, we focus on the charge gap $\Delta_c$. Due to the full symmetry of the blockade structure,
the symmetric sector and the symmetry sector with two charges are degenerate up to perturbative
order $\Om^{10}$, while at twelfth order the term $U_p$ is generated on each plaquette, which leads to
the effective Hamiltonian $ - J \sum_p U_p$. This term
induces the charge gap
\begin{align}
  \Delta_c =4 J =C_{12} \frac{\Omega^{12}}{\Delta^{11}}\,.
\end{align}
The prefactor $C_{12}$ is determined by the sum over the $12!/2^6$ processes
giving rise to the twelfth-order contribution. An efficient
evaluation of the prefactor is possible by dynamic programming, which recursively attributes to the
$3^6-2=727$ different intermediate states the amplitude to reach them from the initial state.
At $U=2 \Delta$, this yields the prefactor
\begin{align}
  C_{12}= 1018657/1728  \approx 589.5 \,,
\end{align}
while for $U\rightarrow \infty$ one obtains $C_{12}=108$ and for $U\rightarrow 0$
one finds the scaling $C_{12}= (63/2)(\Delta/ U)^{5}$.

\subsection{Strong Rabi coupling $\Om \gg\Delta$}

In the opposite limit $\Om \gg \Delta, U$, we only derive the gaps $\Delta_s$
and $\Delta_c$ within first order perturbation theory,
which significantly simplifies the analysis. Then, the ground state energy
is determined by the product state of the dressed states $\ket{-}_s=(\ket{0}_s - \ket{e_{00}}_s)/\sqrt{2}$
on each site with energy $E_0=-2 \Om$ per site of the honeycomb lattice. In first order, we obtain the energy correction
$\Delta/2 + 3 U/16= 7 \Delta/8$ per site from the detuning and the blockade interaction. On the other hand, an excitation
in the symmetric sector is characterized by a single site $s$ in the state
$\ket{+}_s=(\ket{0}_s + \ket{e_{00}}_s)/\sqrt{2}$ with excitation energy $4 \Omega$ and the same diagonal
energy correction as the state $\ket{-}_s$. However, the terms $n_s n_{s'}$ not only provide a diagonal shift but also allow for
a hopping of the excitation along this link with hopping amplitude $U/2 \langle +|n_s|-\rangle\langle -|n_{s'} |+\rangle = U/8$.
With this, the dispersion relation for the excitation
takes the form
\begin{align}
  \epsilon_{\pm}({\bf p})
  = 4 \Om \pm \frac{U}{8}
  \left|1 + e^{i {\bf p}{\bf a}_x}+e^{i {\bf p}{\bf a}_y} \right|
  \,,
\end{align}
which leads to the minimal gap $\Delta_s = 4 \Om - 3 U/8$.

Finally, we focus on the charge gap $\Delta_c$ in the limit $\Om \gg \Delta, U$.
Two charges at adjacent plaquettes are created by the operator $Q^{\alpha}_s$
on one of the two sites of the shared link $\alpha$. At this site, we are
therefore in state $\ket{e_{m_{\alpha}}}_s$ with an energy cost $2 \Omega$ in leading order.
Note that there is a two-fold degeneracy for such adjacent charges.
The repeated application of $Q_{s'}^{\beta}$ along a path on the dual lattice
allows one to separate the two charges. However, this leaves behind a string of sites
in state $\ket{e_{m_{\beta}}}_s$, each with an energy cost $2 \Omega$. Thus
charges are confined with string tension $\sigma = 2 \Omega$. Furthermore,
the degeneracy increases as there are $2^d$ different choices to select the sites along
a path of length $d$. For the charge gap $\Delta_c$, evaluated at distance $d=2$, we
therefore have a 4-fold degeneracy. Then, first order perturbation theory not only provides
a diagonal energy shift $ -\Delta + 3 U/4$ for such charge excitations, but also couples to
two of these different states with $-U/4$. Consequently, the lowest energy state is the equal
superposition of these four configurations and the charge gap reduces to
\begin{equation}
  \Delta_c
  = 4 \Omega -\Delta + \frac{U}{4}
  \,.
\end{equation}

\end{document}